\documentclass[structabstract,usenatbib]{aa}  
\usepackage{graphicx}
\usepackage{txfonts}
\usepackage{natbib}
\begin{document}

\def\la{\lower.5ex\hbox{$\; \buildrel < \over \sim \;$}}
\def\ga{\lower.5ex\hbox{$\; \buildrel > \over \sim \;$}}

 \title{Pinning down the ram-pressure-induced halt of star formation in the Virgo cluster spiral galaxy NGC~4388.}
\subtitle{A joint inversion of spectroscopic and photometric data.}
	 
\author{C. Pappalardo\inst{1}, A. Lan\c{c}on\inst{1}, B. Vollmer\inst{1}, P. Ocvirk\inst{2}, S. Boissier\inst{3}, \& A. Boselli\inst{3} }

 \institute{Observatoire astronomique de Strasbourg, Universit\'e de Strasbourg \& CNRS UMR7550, 11 rue de l'Universit\'e, 67000 Strasbourg, France 
  		\email{ciro.pappalardo@astro.u-strasbg.fr}
         \and
             Astrophysikalisches Institut Potsdam, An der Sternwarte 16, 14482 Potsdam, Germany 
             \and
             Laboratoire d'Astrophysique de Marseille, OAMP, Universit\'e Aix-Marseille \& CNRS UMR6110, 38 rue Fr\'ed\'eric Joliot Curie, 13388 Marseille Cedex 13, France 
             }
\titlerunning{The stripping age of NGC~4388}
\authorrunning{Pappalardo et al.}

   \abstract
   {In a galaxy cluster, the evolution of spiral galaxies depends on their cluster environment. Ram pressure due to the rapid motion of a spiral galaxy within the hot intracluster medium removes the galaxy's interstellar medium from the outer disk. Once the gas has left the disk, star formation stops. The passive evolution of the stellar populations should be detectable in optical spectroscopy and multi-wavelength photometry.} 
   { The goal of our study is to recover the stripping age of the Virgo spiral galaxy NGC~4388, i.e. the time elapsed since the halt of star formation in the outer galactic disk using a combined analysis of optical spectra and photometry.}
   {We performed VLT FORS2 long-slit spectroscopy of the inner star-forming and outer gas-free disk of NGC~4388. We developed a non-parametric inversion tool that allows us to reconstruct the star formation history of a galaxy from spectroscopy and photometry. The tool was tested on a series of mock data using Monte Carlo simulations. The results from the non-parametric inversion were refined by applying a parametric inversion method.}
   {The star formation history of the unperturbed galactic disk is flat. The non-parametric method yields a rapid decline of star formation $\la 200$~Myr ago in the outer disk. The parametric method is not able to distinguish between an instantaneous and a long-lasting star formation truncation. The time since the star formation has dropped by a factor of two from its pre-stripping value is $190 \pm 30$~Myr.} 
   {We are able to give a precise stripping age that is consistent with revised dynamical models.}


   \maketitle

%

%
 \section{Introduction}

Depending on the environment in which they move, spiral galaxies undergo different processes that can modify their structure significantly (see \citealt{bos} and references therein):
\begin{itemize}
\item[-] gravitational effects (e.g. tidal interactions in galaxy-galaxy encounters),
\item[-] hydrodynamical effects (e.g. ram pressure stripping or thermal evaporation),
\item[-] hybrid processes, i.e. those involving both types of effects, such as preprocessing and starvation.
\end{itemize}
The closest cluster of galaxies in the northern hemisphere is the Virgo cluster ($d\approx 16.7$ Mpc, \citealt{ya}) with a mass of $M = 1.2 \times 10^{15} M_{\odot}$ and a radius of about 2.2 Mpc (\citealt{fo}). 
Virgo is an evolving cluster that is still dynamically active. The cluster-core is centered on M87, the most massive elliptical galaxy. Other subclumps are falling into the potential well of the cluster.

One important characteristic of Virgo spiral galaxies is their lack of gas (\citealt{gi}, \citealt{ch}). The amount of atomic gas in Virgo spirals is up to 80\% less than for field galaxies of the same size and morphological type. Virgo spirals show truncated HI disks (\citealt{gi}, \citealt{cay}) with respect to their optical disks. The galaxies on radial orbits are on average more HI deficient than the ones on circular orbits (\citealt{dr2}). \cite{chu} find long HI tails associated with spiral galaxies located at distances from 0.5 to 1 Mpc from the cluster center. For these cases, ram pressure stripping is the most probable cause.
The interstellar medium (ISM) of a spiral galaxy that is moving inside the potential well of a cluster, undergoes pressure due to the intracluster medium (ICM) that is hot ($T_{\rm{ICM}} \approx 10^7 - 10^8 $ K) and tenuous ( $\rho_{\rm{ICM}} \approx 10^{-3} - 10^{-4}$ atoms cm$^{-3}$). If this pressure is higher than the restoring force due to the galactic potential, the galaxy loses gas from the outer disk. Quantitatively this is expressed by the \cite{gu} criterion:
\begin{equation}
\rho_{\rm{ICM}} v_{\rm{gal}}^2 \ge 2 \pi G \Sigma_{\rm{star}}\Sigma_{\rm{gas}},
\end{equation}
where $\rho_{\rm{ICM}}$ is the density of the ICM, $v_{\rm{gal}}$ the peculiar velocity of the galaxy inside the cluster, and $\Sigma_{\rm{star}}$ and $\Sigma_{\rm{gas}}$ are the surface density of stars and gas, respectively.
Ram pressure stripping has been studied theoretically (e.g. \citealt{vol}, \citealt{sh}, \citealt{qu}, \citealt{ab}, \citealt{ro}) and observationally (e.g. \citealt{ke}, \citealt{so}, \citealt{cay}, \citealt{wa}, \citealt{chu}). 

NGC 4388 is a highly inclined Seyfert 2 spiral galaxy (of type Sab) with $m_B=12.2$ mag and a radial velocity of $V_{\rm{rad}} \sim 1400$ km s$^{-1}$ with respect to the cluster mean. It is located at a projected distance of 1.3$^\circ$ ($\approx$ 400 kpc at 16.7\,Mpc distance) from the Virgo cluster center (M87). NGC 4388 has lost 85\% of its HI gas mass (\citealt{cay}). The HI distribution is strongly truncated within the optical disk. By observing the galaxy in H$\alpha$ \cite{vi2} found a large plume of ionized gas extending 4 kpc above the plane. In subsequent SUBARU observations \cite{yo} reveal a more extended tail out to 35 kpc to the northeast. \cite{vo} performed 21-cm line observations with the Effelsberg 100-m radio telescope. They 
discovered neutral gas associated with the H$\alpha$ plume out to at least 20 kpc NE of NGC 4388's disk, with an HI mass of $6 \times 10^7$ M$_{\odot}$. 
With interferometic observations  \cite{oo} show that this HI tail is even
more extended, with a size of $110\times 25$ kpc and a mass of $3.4 \times 10^8$ M$_{\odot}$.

In the case of NGC 4388, \cite{vo} estimate that ram pressure stripping is able to remove more than 80\% of the galaxy's ISM, consistently with the observations of \cite{cay}. They also estimate that the galaxy passed the cluster center $\sim 120$ Myr ago. However this estimation is based on the observations of \cite{vo}. The more extended HI tail found by \cite{oo} implies that this timescale increases.

Once the gas has left the galactic disk, star formation, which is ultimately fueled by the neutral hydrogen, stops. Stripping is believed 
to progress inwards from the outermost disk on timescales of $10^8$ years, but at any given radius, stripping happens on a shorter timescale. 
The stellar populations 
then evolve passively, and this is detectable both in optical spectra and in 
the photometry. The detailed analysis of the stellar light provides essential 
information on the stripping age, i.e.the time elapsed since the halt of 
star formation. These constraints are independent of dynamical models and 
can be compared to those derived from the gas morphology and kinematics. 
Crowl \& Kenney (2008) analyze the stellar populations of the gas-free 
outer disk of NGC 4388, 5.5 kpc off the center (with our adopted distance 
of 16.7 Mpc). They used SparsePak spectroscopy and GALEX photometry. Their 
age diagnostics are based on Lick indices and UV fluxes, which they 
compare to stellar population synthesis predictions (models of Martins 
et al. 2005). Assuming that the past star formation rate in the galaxy 
has been constant on long timescales, they conclude the stripping of 
the gas from the outer disk occurs $225 \pm 100$ Myr ago.
Our renewed analysis is based on VLT FORS2 long-slit spectroscopy of NGC 4388 taken at two positions:
\begin{itemize}
\item The first slit was pointed at 1.5 kpc from the center, in a gas-normal star-forming region of the disk. We did not take the exact center to avoid bulge contamination.
\item The second slit was pointed at 4.5 kpc from the center, just outside the star-forming gas disk.
\end{itemize}
The spectra were combined with multi-wavelength photometry. 
The long-term star formation history of the galaxy was obtained using 
an extension of the non parametric inversion method of Ocvirk et al. (2006). 
It combined the information provided by the spectroscopic and photometric data to determine the star formation history using minimal constraints on the 
solutions. The results were then refined using a parametric method that 
assumes the halt of star formation at the observed outer position in the 
disk occurs quasi-instantaneously.

The paper is structured as follows. In Sect.s \ref{observations} and \ref{photometry} we describe the observations and explain how we extract the photometry that we use as input data. In Sect. \ref{method} we introduce the new approach used in this paper that combines spectral and photometric analysis. In Sect. \ref{results} and \ref{discussion} we explain the results of this new method in the case of NGC 4388. Finally, in Sect. \ref{conclusion} we give our conclusions and compare our results to previous work.

\section{Observations}
\label{observations}

\subsection{Data set}

We performed observations of NGC4388 on May 2, 2006 at the European Southern Observatory Very Large Telescope (VLT) facility on Cerro Paranal, Chile.
The instrument we used was the FOcal Reducer and low dispersion Spectrograph 2 (FORS2, \citealt{app}) in long-slit spectroscopy observing mode.
The detector system consisted of two 4096 $\times$ 2048 CCD. 
We chose 2 $\times$ 2 binning of the pixels (image-scale 0.252$''$/ binned pixel) and high-gain readout.
We selected  grism GRIS-600B with a wavelength range of $3350 - 6330$  $\AA$ and a resolution of 1.48 $\AA$ / binned pixel. The data were acquired through a $1''$ slit, yielding a resolving power of $\lambda/ \Delta \lambda \approx 780$ at the central wavelength.

\begin{figure}
   \centering
    \resizebox{\hsize}{!}{\includegraphics{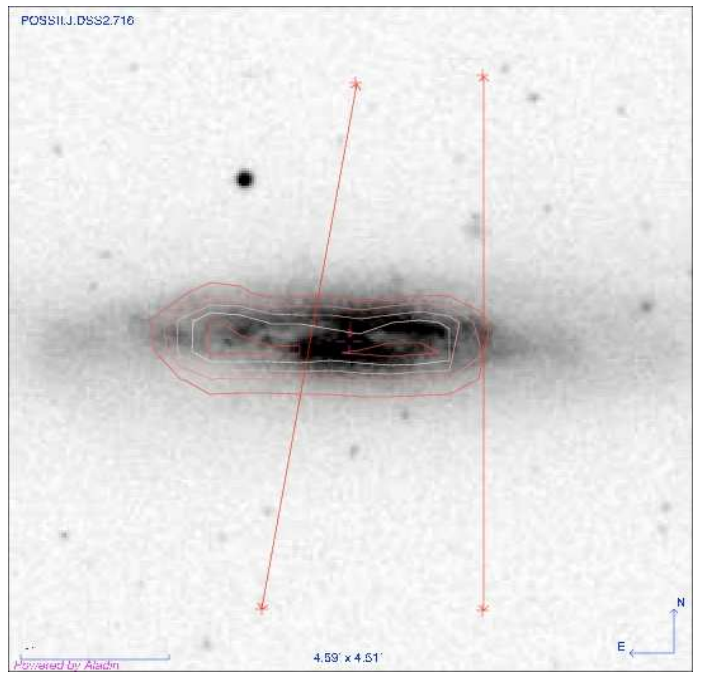}}
      \caption{B-band POSSII image of NGC4388 with HI contours. The slit positions for the inner and the outer regions are overlaid.}
         \label{slits}
   \end{figure}

The two slit positions used for the acquisition are shown in Fig.~\ref{slits}
together with HI contours. Individual exposure times were 600\,s for the inner region and 1350\,s for the outer.
The total integration time for each region were 1800\,s and 9450\,s (Table \ref{obs}). 
We used LTT 4816 and LTT 7987, two white dwarfs with $m_V=13.96$ and $m_V=12.23$, respectively (\citealt{ba}, \citealt{pe}) as spectrophotometric standards.

\begin{table*}
\caption{Journal of observations (ESO program ID 77.B-0039(A)). Seeing, airmass, and UT (Universal Time) refer to the beginning of each acquisition.}             
\label{obs}      
\centering                          
\begin{tabular}{c c c c c c c}        
\hline\hline                 
Observation        &    Frame &Exp. Time (sec) & Seeing   & Airmass & Date              & UT \\ 
\hline                     
LTT4816 Ph.St.  &           1       &35                      & 0.91$''$      &     1.484            & 2006 May 2 & 23:06\\
NGC4388 Inner &            1       &600                      & 0.9$''$      &     1.688            & 2006 May 2 & 23:45\\
NGC4388 Inner &            2       &600                      & 1.11$''$    &     1.616            & 2006 May 2 & 23:56\\
NGC4388 Inner &            3       &600                      & 1.10$''$    &     1.554            & 2006 May 3 & 00:07\\
NGC4388 Outer&            1       &1350                    &  0.89$''$   &     1.330            & 2006 May 3 & 01:11\\
NGC4388 Outer&            2       &1350                    &  0.83$''$   &     1.289            & 2006 May 3 & 01:34\\
NGC4388 Outer&            3       &1350                    &  0.96$''$   &     1.264            & 2006 May 3 & 01:58\\
NGC4388 Outer&            4       &1350                    &  0.67$''$   &     1.256            & 2006 May 3 & 02:21\\
NGC4388 Outer&            5       &1350                    &  0.61$''$   &     1.423            & 2006 May 3 & 04:10\\
NGC4388 Outer&            6       &1350                    &  0.55$''$   &     1.519            & 2006 May 3 & 04:33\\
NGC4388 Outer&            7       &1350                    &  0.55$''$   &     1.651            & 2006 May 3 & 04:57\\
LTT7987 Ph.St.  &           1       &40                          & 1.34$''$   &     1.010            & 2006 May 3 & 10:35\\
\hline                                   
\end{tabular}
\end{table*}

\subsection{Data reduction}
\label{secdatareduction}

Data were reduced with the {\tt{images}} and {\tt{noao.twodspec}} packages of the Image Reduction and Analysis Facility (IRAF)\footnote{IRAF is distributed by the National Optical Astronomy Observatory, which is operated by the Association of Universities for Research in Astronomy, Inc., under cooperative agreement with the National Science Foundation.} \citep{to}.
We created average bias and flat-field images using daytime calibration, and we removed cosmic ray hits using a $3\sigma$-clipping rejection.

After bias and flat-field correction, we combined the individual galaxy exposures and applied the wavelength calibration to the spectral axis of the images.
The aperture used for extracting the 1D spectra covers a length of $32.76''$ along the slit for both the inner and the outer regions of NGC 4388.
The sky subtraction was based on a linear fit to the sky windows on either side of the aperture.
Finally, we used the spectrum of LTT 4816 and its intrinsic fluxes \citep{ha} 
to calibrate the flux of the spectra.
The flux calibration is $\approx 20 \%$ accurate, based on the comparison of the energy distributions obtained using either LTT4816 or LTT7987  as a standard.
 
The final reduced spectra for the inner and outer regions are shown in Sect.~\ref{results}.
The average  signal-to-noise ratio per pixel for the inner and outer regions are $\approx 60$ and $\approx 26$, respectively, inside the wavelength ranges used in the analysis. 

\section{Photometry}
\label{photometry}

For the photometry we used the following archive data:

\begin{itemize}
 \item GALEX (\citealt{gil}): FUV and NUV.
 \item SDSS (Rel. 6, \citealt{ad}):  $u' , g' , r' , i',$ and $z'$.
 \item  2MASS (\citealt{sk}) : $J$, $H,$ and $K$. 
\end{itemize}

The archive images were resampled onto the coordinate grid used in FORS2 observations using ALADIN (\citealt{bo}).
The photometric zero points were derived using the star 2MASS $J122549.86+124047.9$ (Table \ref{mag}).
\begin{table*}
\caption{NGC 4388 photometry. $1^{st} - 2^{nd}$ row: magnitude and error for the reference star 2MASS $J122549.86+124047.9$. $3^{rd} - 4^{th}$ row:  magnitude and error for the inner region. $5^{th} - 6^{th}$ row : magnitude and error for the outer regions. $7^{th} - 8^{th}$ row: total uncertainty for inner and outer regions. For the SDSS and 2MASS filters the magnitude are given in AB and Vega systems, respectively.}
\label{mag}
\centering
\begin{tabular}{|c|c|c|c|c|c|c|c|c|c|c|}
\hline
 Filter &  $FUV$  &$NUV$  & $u'$   & $g'$   & $r'$    & $i'$    & $z'$  & $J$    & $H$  & $K$  \\
\hline
$m$   &    $-$ &  $-$   &16.78 &15.06 &14.35 &14.69 &14.07 &13.08 &12.65 &12.52\\
$\sigma_{star}$        & $-$ & $-$ & 0.025   & 0.034 & 0.017  & 0.002 & 0.016  & 0.021 & 0.03    & 0.024  \\
Inner &   20.40     & 19.95   &18.45 &17.10 &16.37 &16.03 &15.76 &14.38 &13.60 &13.32\\
$\sigma_{SLIT}$-Inner &0.17& 0.04&0.01    &0.02   &0.04   &0.04   &0.04    &0.04   &0.02    &0.05  \\
Outer & 22.21      &  21.30  &19.49 &18.07 &17.47 &17.18 &17.03 &16.08 &15.06 &14.75 \\
$\sigma_{SLIT}$-Outer&0.31&0.07&0.08&0.06   &0.07    &0.07   &0.03   &0.33   &0.04    &0.12   \\
\hline
\hline
$\sigma_{TOT}$-Inner &0.2 & 0.1 &0.03  &0.04   &0.04   &0.04   &0.04   &0.05   &0.04    &0.06  \\
\hline
$\sigma_{TOT}$-Outer &0.33& 0.12&0.07  &0.07   &0.07   &0.07   &0.03   &0.33   &0.05    &0.12  \\
\hline
\end{tabular}
\end{table*}
For the GALEX images, we used the formulae of \cite{mo} to convert counts per second ($CPS$) into AB magnitudes.
The defined zero points are accurate to within $\pm 10 \%$ (\citealt{mo}). 

The exact locations of the two spectroscopic apertures on the images were determined by comparing the wavelength-averaged profile along the FORS2 slits with cuts through the SDSS $g'$ image. In both cases a clean peak in the cross correlation function allowed us to identify the slit position to within 1 pixel size.
The GALEX images have a spatial resolution of about 4\arcsec. Light 
measured within a 1\arcsec\ aperture is contaminated by neighboring areas,
but this effect is less than the uncertainties already accounted for,
which are large because of low signal levels.

We estimated the uncertainties on the final magnitudes for the galaxy apertures from a combination of the uncertainties on the flux measurements of the reference star (zero point) and those of the positioning of the slit (see Table \ref{mag}):

\begin{equation}
\sigma_{\rm{TOT}}^2 = \sigma_{\rm{SLIT}}^2 + \sigma_{\rm{STAR}}^2.
\end{equation}

\section{Method}
\label{method}

The spectral energy distribution (SED) of a galaxy can be considered as the integrated light produced by different stellar populations with different ages and metallicities. 
The light and mass contributions of each population depend on the star formation history of the observed region during the galaxy's life.

If we define an initial mass function and a set of isochrones, we can obtain the intrinsic spectrum $B^0(\lambda,t,Z)$ of the single stellar population of age $t$, metallicity $Z$, and unit mass by integrating over the stellar masses.
Assuming that the metallicities of the stars can be described by a single-valued age-metallicity relation $Z(t)$, we can derive the unobscured SED of a galaxy at rest:

\begin{equation}
F_{\rm{rest}}(\lambda) = \int_{t_{\rm{min}}}^{t_{\rm{max}}}SFR(t) \ B^0(\lambda,t,Z(t))\ \mathrm{d}t,
\label{frest}
\end{equation}
\noindent
in which $SFR(t)$ represents the mass of new stars born per unit time, with the convention that $t_{\rm{min}} = 0$ is today and $t_{\rm{max}}$ the Hubble time.

Since we observe the light and not the mass in a galaxy, it is more convenient to convert the mass-weighted spectral basis $B^0(\lambda,t,Z)$ into a luminosity-weighted basis.
The difference is that  $B^0(\lambda,t,Z)$ defines the spectrum of a single stellar population (SSP) of unit mass, and $B(\lambda,t,Z)$ defines an SSP spectrum of unit flux.
Instead of mass contributions we deal with light contributions, thus converting the SFR into the luminosity-weighted stellar age distribution (hereafter SAD):

\begin{equation}
\Lambda(t) = \frac{SFR(t)}{\Delta \lambda} \int^{\lambda_{\rm{max}}}_{\lambda_{\rm{min}}}B^0(\lambda,t,Z) \ \mathrm{d}\lambda,
\end{equation}
\noindent
where $\Delta \lambda = \lambda_{\rm{max}} - \lambda_{\rm{min}}$ is the available wavelength domain, and $\Lambda(t)$ gives the contribution to the total light from the stars of age $[t,t+\mathrm{d}t]$.
\noindent
With

\begin{equation}
B(\lambda,t,Z)=\frac{B^0(\lambda,t,Z)}{\frac{1}{\Delta \lambda} \int^{\lambda_{\rm{max}}}_{\lambda_{\rm{min}}}B^0(\lambda,t,Z)\ \mathrm{d} \lambda},
\end{equation}
\noindent
Eq.~\ref{frest} becomes

\begin{equation}
F_{\rm{rest}}(\lambda) = \int^{t_{\rm{max}}}_{t_{\rm{min}}}\Lambda(t) B(\lambda,t,Z(t)) \ \mathrm{d}t.
\label{frest2}
\end{equation}
\noindent
In the same way we can associate a photometric value with each spectrum $B(\lambda,t,Z(t))$ for the set of bandpasses defined in Table \ref{mag}:

\begin{equation}
B_{\rm{phot}}( y,t,Z(t)) = \frac{\int_{\lambda_{\rm{min}}}^{\lambda_{\rm{max}}} B(\lambda,t,Z(t)) \cdot T_y(\lambda) \ \lambda \ \mathrm{d}\lambda}{\int_{\lambda_{\rm{min}}}^{\lambda_{\rm{max}}} T_y(\lambda) \ \lambda \ \mathrm{d}\lambda},
\end{equation}
\noindent
in which $y = [ FUV , NUV , u' , g' , i' , r' , z' , J , H , K]$, and $T_y$ is the transmission curve associated with each $y$. 
Unobscured photometry $F_{\rm{phot}}(y)$ is given by

\begin{equation}
\begin{array}{rl}
F_{\rm{phot}}(y) & = \int^{t_{\rm{max}}}_{t_{\rm{min}}} SFR(t) \, B^0_{\rm{phot}}(y,t,Z(t))\ \mathrm{d}t \\
           & = \int^{t_{\rm{max}}}_{t_{\rm{min}}} \Lambda(t) \, B_{\rm{phot}}(y,t,Z(t))\ \mathrm{d}t.
\end{array}
\label{fphoto}
\end{equation}
\noindent
Eq.~\ref{frest2} is not complete, because it does not take other effects into account that can modify the final shape of the SED of a galaxy, as follows:

\smallskip

$\bullet$ Extinction: when fitting spectroscopic data we use a flexible continuum correction 
that can account both for the reddening due to dust and for flux calibration errors. 
The method allows us to use the information present in the spectral lines without using the continuum of the spectrum,
which is preferable when the flux calibration is not perfect. 
In practice, we define a set of equally spaced anchor points across the wavelength range of the optical spectra. 
Their ordinates are then optimized in such a way that the cubic spline interpolation through the points cancels out any SED difference
between models and data. We refer to this adjustable correction as NPEC hereafter (for non parametric estimate of the continuum).

For the photometry, however, extinction must be accounted for explicitly. We adopted the attenuation law of \cite{cal} that uses the colour excess 
$E(B-V)$ as a single parameter (see, however, Sect.\,\ref{discussion}).

$\bullet$ Radial velocities of the stars: we assumed that the velocities of stars of all ages along the line of sight have the same velocity distribution. We can approximate the spectrum of a galaxy as the convolution of the spectrum at rest and a line of sight velocity distribution $g(v)$:

\begin{equation}
\phi(\lambda) = \int_{v_{min}}^{v_{max}}F_{\rm{rest}} \Big( \frac{\lambda}{1+v/c} \Big) \cdot  g(v) \frac{\mathrm{d}v}{1+v/c},
\label{los}
\end{equation} 
where $F_{\rm{rest}}$ is defined in Eq.~\ref{frest2}, and $c$ is the velocity of light.
The integrals in Eqs.\,\ref{frest2}, \ref{fphoto}, and \ref{los} are transformed into discrete sums for numerical treatment.

Recovering the star formation history ultimately means finding a $\Lambda(t)$ that fulfills Eqs. \ref{frest2} and \ref{fphoto},
while accounting for the corrections just described. To do this, we used two different approaches:

\begin{itemize}

\item a non parametric method in which we recover the star formation history by resolving the associated inverse problem with regularized methods (see \citealt{oc});

\item a parametric method in which we define a set of possible $\Lambda(t)$ depending on one free parameter, the time
elapsed since the stripping event.  We then find the most probable solution by minimizing the classical $\chi^2$ function.

\end{itemize}

\smallskip
\noindent
The non parametric method has the advantage of providing the star formation history and trends in the age-metallicity relation of the galaxy with minimal constraints on their shape. On the other hand, the regularization of the problem did not allow us to recover functional forms with large gradients, such as those expected for a ram pressure stripping event. This is why we combined the results from the non-parametric analysis with a parametric analysis.

\smallskip

We chose as SSP library the models of \cite{bc03} that cover  a time interval $\Delta t = [0 - 19.5]$ Gyr, a wavelength range $\lambda \lambda = [100 - 24850] \AA$ with an average spectral resolution (FWHM) $R = 2000$ at optical wavelengths. These SSP spectra are constructed with stellar spectra from  \cite{leb}. The underlying stellar evolution tracks are those of \cite{al}, \cite{br}, \cite{fa}, \cite{fa2} and \cite{gir}.

\subsection{Non parametric method}
\label{npmethod}

The non parametric method used is a modification of STECKMAP\footnote{http://astro.u-strasbg.fr/$\sim$ocvirk/} (\citealt{oc}, \citealt{oc2}), extended to deal with spectroscopy and photometry jointly. Assuming gaussian noise in the data, we estimate the most likely solution by minimizing the following $Q_\mu (\mathbf{X})$ function for the star formation history and age-metallicity relation:

\begin{equation}
Q_\mu(\mathbf{X}) = ( 1 - \alpha) \cdot \chi^2_{\rm{spec}}(\mathbf{X}) + \alpha \cdot \chi^2_{\rm{phot}}(\mathbf{X}) + \mu \cdot P(\mathbf{X}),
\label{qfun}
\end{equation}
\noindent
in which

\begin{itemize}
\item $\mathbf{X}$ is a vector that includes the SAD, the metallicity evolution, the color excess $E(B-V)$ for the photometry, the NPEC correction vector for spectral analysis and a spectral broadening function (BF). The last includes the effects of the line of sight velocity distribution and the broadening introduced by the instrumentation.

\smallskip

\item 

\begin{equation}
\chi^2_{\rm{spec}}(\mathbf{X}) = \frac{1}{N_{\lambda}} \sum_{i=1}^{N_\lambda}
\frac{\big(F_{\rm{mdlspec}}(\lambda) - F_{\rm{spec}}(\lambda)\big)^2}{\sigma_{\rm{spec}}^2},
\end{equation}

is the $\chi^2$ associated with the spectrum.
$F_{\rm{mdlspec}}(\lambda)$ is the model with a defined $\mathbf{X}$ and $F_{\rm{spec}}(\lambda)$ is the observed spectrum, $\sigma_{\rm{spec}}^2$ is the associated error in the observations, and $N_{\lambda} \sim 1700$ is the number of points in the spectrum, which we can consider approximately equal to the number of degrees of freedom of the problem.

\smallskip

\item

\begin{equation}
\chi^2_{\rm{phot}}(\mathbf{X}) = \frac{1}{N_y} \sum_{i=1}^{N_y}
\frac{\big(F_{\rm{mdlphot}}(y) - F_{\rm{phot}}(y)\big)^2}{\sigma_{\rm{phot}}^2},
\end{equation}

is the $\chi^2$ associated with the photometry, where $F_{\rm{mdlphot}}(y)$ is the model photometry obtained from Eq.~\ref{fphoto} with a chosen $\mathbf{X}$, $F_{\rm{phot}}(y)$ are the photometric data and $\sigma_{\rm{phot}}^2$ are the errors in the measures, and $N_y$ represents the number of bandpasses used.

\medskip

\item 

\begin{equation}
\mu P(\mathbf{X}) =  \mu_xP(\mathrm{SAD}) + \mu_Z P(\mathrm{AMR}) 
\end{equation}
\smallskip

\hspace*{1.5cm} $+ \mu_{NPEC} P(\mathrm{E}) + \mu_{BF}P(\mathrm{BF}) $

\smallskip
\noindent
is a penalty function needed to regularize the problem. As shown in \cite{oc}, the inverse problem associated with Eqs.~\ref{frest2} and \ref{fphoto} is ill-posed. The function $P(\mathbf{X})$ is chosen to yield high values when the SAD and AMR are a very irregular function of time or when the BF or the NPEC correction are too chaotic. The set of $\mu = (\mu_x,\mu_Z,\mu_{NPEC},\mu_{BF})$ are adjustable parameters that control the weight of each $P(\mathbf{X})$ in the final estimation.

\item $\alpha$ determines the relative weights of the photometric and spectroscopic constraints.

\end{itemize}

\subsection{Parametric method}
\label{pmethod}

In this case we assume an exponential star formation history
before the stripping event, a metallicity and a BF, all consistent 
with the non parametric results. We reproduce the stripping by a linear decrease in the SFR on timescales between $0 \le t \le 100$~Myr. For each of these truncated star formation histories, we calculate the NPEC correction and $E(B-V)$ that produce the lowest $\chi^2$.
The minimum of $\chi^2(t)$ is taken to provide the most likely stripping age, which we define as the time elapsed since star formation has dropped by a factor of two from its pre-stripping value. The spectroscopic and photometric contributions to the $\chi^2$ are weighted as before:

\begin{equation}
\chi^2_{\rm{tot}}(\mathbf{X},t) = (1-\alpha) \cdot \chi^2_{\rm{spec}}(\mathbf{X},t) + \alpha \cdot \chi^2_{\rm{phot}}(\mathbf{X},t),
\label{ptot}
\end{equation}
where $\alpha$, $\chi^2_{\rm{spec}}(\mathbf{X},t)$, and $\chi^2_{\rm{phot}}(\mathbf{X},t)$ have the same meaning as in Sect. \ref{npmethod}.

\section{Results}
\label{results}

\subsection{Non parametric Inversion}
\label{nonpmethod}
First we point out that the weightings $\mu$ for the penalty are a central issue of the non parametric method and that their determination is not a trivial problem. There are different ways of fixing these values \citep{ti}. In the case of nonlinear problems, one has to proceed via empirical tuning \citep{cb} to define the values of $\mu$ below which the results present artifacts, and above which the smoothness of the solution is completely due to the penalty.

We chose as penalty functions two finite difference operators ($\mathbf{L = D_1}$ and 
$\mathbf{L = D_2}$ in \citealt{oc}) that favor small first derivatives of the AMR and small second derivatives of the SAD and BF. The penalties for the star formation history and the metallicity evolution are defined in logarithmic timescale.  
We explored the effect of the weight coefficients $\mu$ through a campaign of inversions of artificial spectra.
For the NPEC correction, we found it was not necessary to require smoothness explicitely, 
and the penalization simply acts to normalize the continuum correction 
(thus avoiding the degeneracy between the absolute values of this correction and the star formation rates).
The method recovers similar well-behaved BF for a wide range of $\mu_{BF}$. The campaign showed that it is 
not possible to recover more than a tentative 
linear trend in the AMR with the data available to us, and we chose to
simplify the problem with constant metallicity (i.e. large $\mu_Z$). 
A realistic metallicity evolution shows a rapid increase at long lookback times and little increase over the past 5~Gyr (\citealt{bp}). From our method we derived a time-bin averaged metallicity over a Hubble time $\langle Z \rangle$. For typical models of \cite{bp}, this mean metallicity is $\Delta Z \sim 0.005$-$0.01$ lower than the time-bin averaged metallicity over the past 5~Gyr. 

For the stellar age distribution, we adopt the smallest $\mu_x$ that provides robust solutions, i.e. low sensitivity to the noise
in the data. 
This choice was based on successive inversions of pseudo-data with different $\mu_x$. The stability of the results has been tested via Monte Carlo simulations: we added a Gaussian noise with $\sigma$ equal to the noise of the data to the input spectrum and repeated the inversion. In this way we fixed the limits for the $\mu_x$ value. In the campaign of tests, we also verified that, within reasonable limits, the shape of the initial 
guess does not affect the recovered solution significantly. We ended up taking a semi-analytical model of \cite{bp} as an
initial guess for the star formation rate and the age-metallicity relation, and a constant for the BF and the 
NPEC correction. 

For each region we reconstructed the star formation history using the VLT spectrum alone ($\alpha = 0$), the photometry alone ($\alpha = 1$) and spectrum/photometry at the same time ($\alpha = 0.5$). In the combined analysis the choice of $\alpha = 0.5$ gives, in the final estimation of $Q(\mathbf{X})$, the same weight to the spectral and photometric analysis. Our conclusions are not significantly affected by the choice of $\alpha$, as verified through a campaign of inversions.

\subsubsection{Inner Region}
\begin{itemize}
\item VLT spectrum ($\alpha = 0)$.

The spectrum shows a fit (top panel in Fig.~\ref{innerfit}) with $\chi^2_{\rm{spec}} = 2.2$ and a star formation history (Fig.~\ref{innersfr01}) that is approximately flat.  
The time-bin averaged metallicity is nearly solar, $\langle Z \rangle = 0.018 \pm 0.003$. For comparison we took three emission lines ([OII]3727, H$\beta$4861, [OIII]5007) to recover the metallicity using the strong line method (\citealt{pi}). We obtained $\langle Z \rangle \approx 0.021$ consistent with the results of the inversion.
 The broadening function is centered on -100 km/s. The shift and the width of the broadening function are consistent with expectation based on galactic rotation and non-circular motions in a barred potential (\citealt{vi2}).
   \begin{figure*}
   \centering
    \includegraphics[angle = -90,width = 0.9\textwidth]{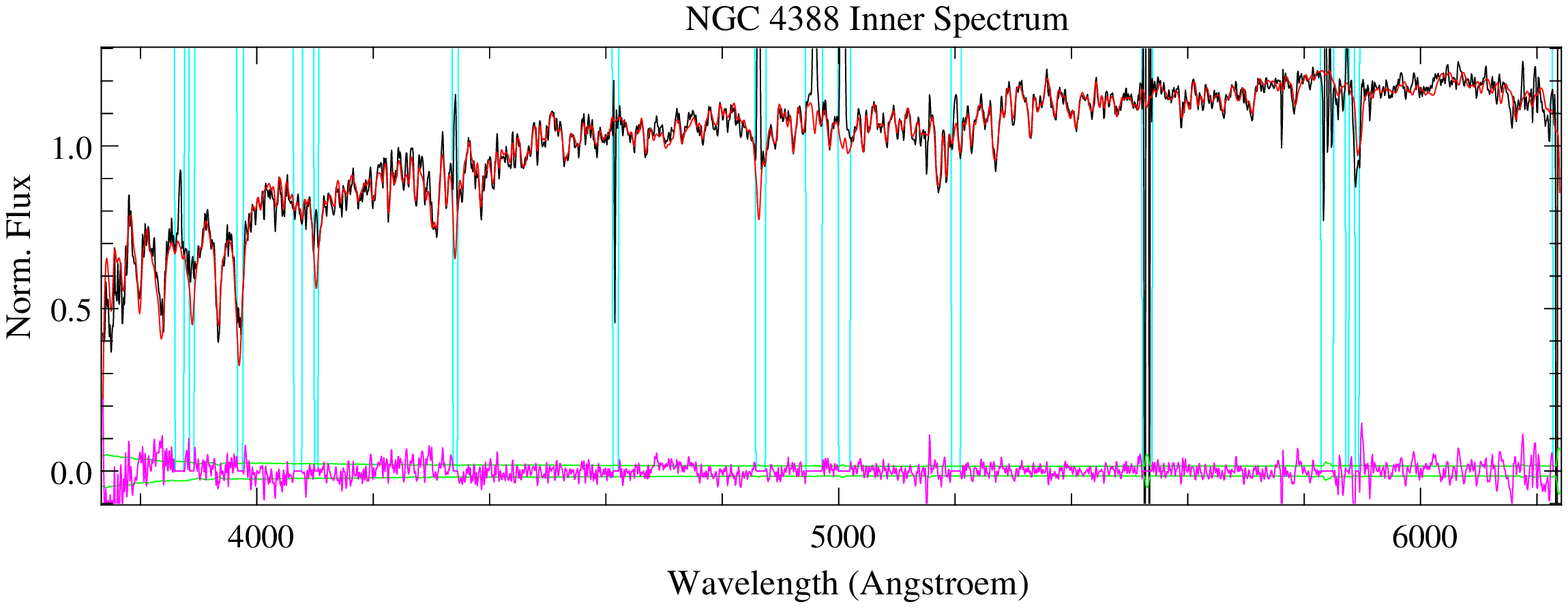}
    \includegraphics[angle = -90,width = 0.9\textwidth]{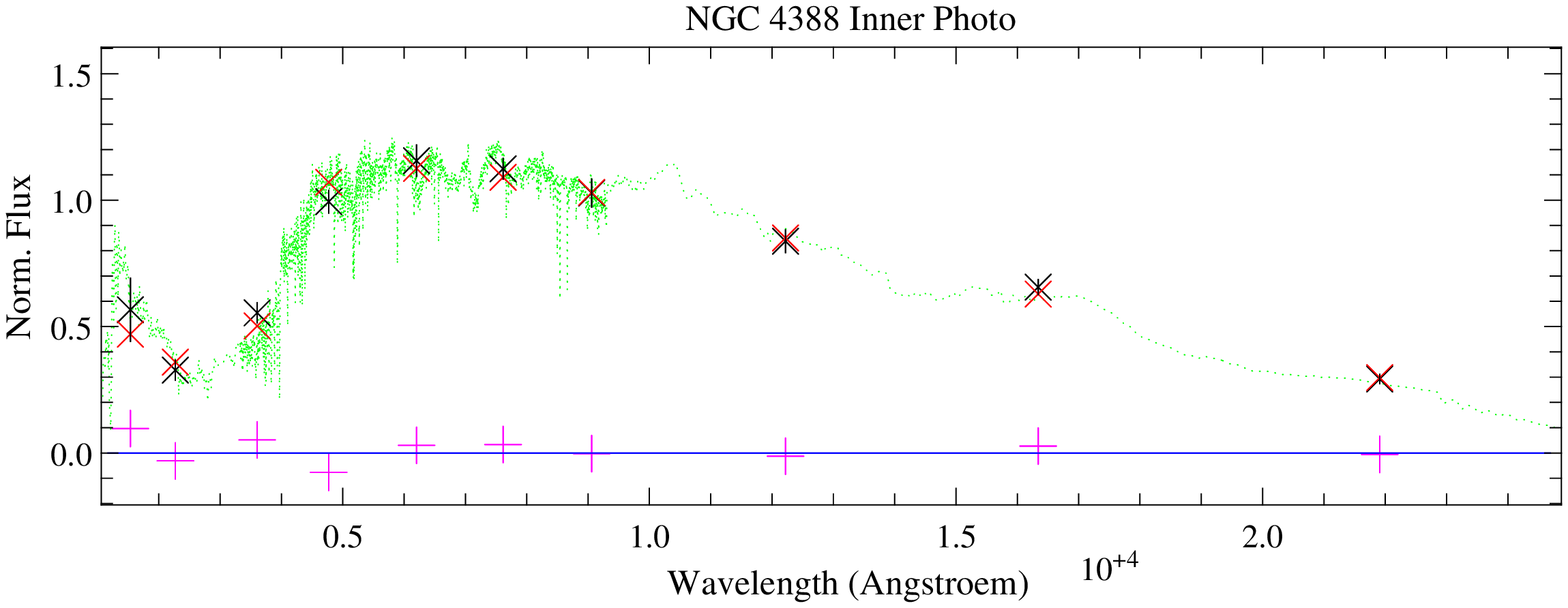}
    \includegraphics[angle = -90,width = 0.9\textwidth]{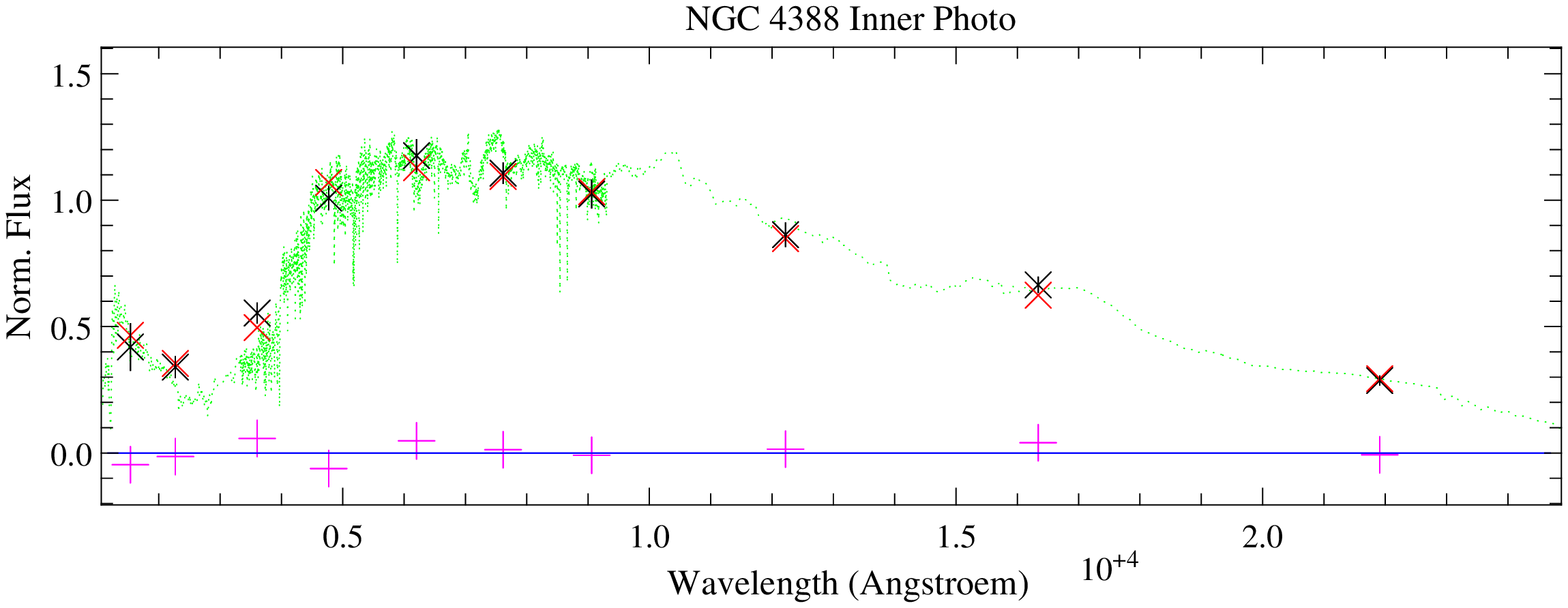}
   \caption{NGC 4388 inner region fits. Top panel: spectra (black line) and best fit (red line) obtained from inversion with $\alpha = 0$. The bottom of the panel shows the residuals (magenta line) and the $\sigma$ (green line).  The cyan vertical lines show the mask used for emission and sky lines. Middle panel: photometry (black crosses) and best fit (red crosses) obtained from inversion with $\alpha  = 1$ overplotted to the corresponding flux (green dotted line). In the bottom of the panel are shown the residuals (magenta crosses). Bottom panel: same as the middle panel, but using $\alpha = 0.5$. We do not show the fit of the spectrum for the case $\alpha = 0.5$, because it is 
   indistinguishable by eye from the top panel.} 
   \label{innerfit}
   \end{figure*}

\begin{figure}
\centering
\includegraphics[,width = 0.45\textwidth]{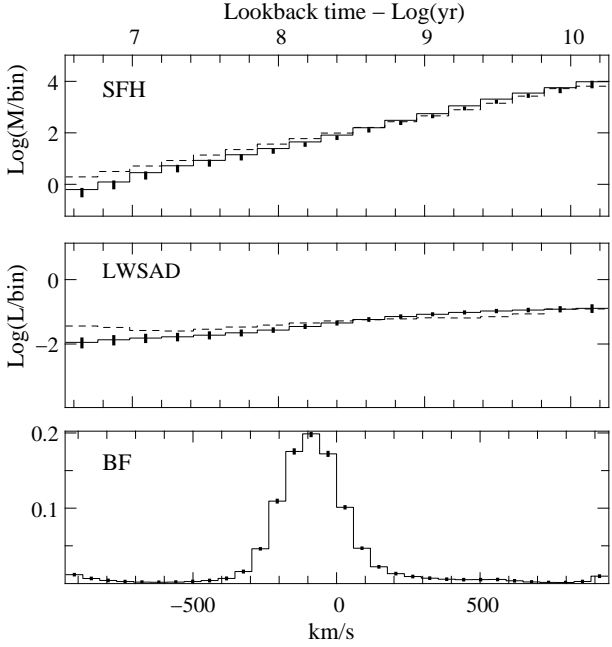}
\caption{Non parametric spectral inversion ($\alpha$ = 0) of the inner region of NGC 4388. Top and middle panel: star formation history (SFH) and luminosity weighted stellar age distribution (LWSAD) vs. lookback time. Bottom panel: spectral broadening function (BF). In each panel the solid lines shows the results of the inversion with associated error bars obtained from 30 Monte Carlo simulations. The dashed line shows a constant star formation rate. All the figures are in arbitrary units.}
\label{innersfr01}
\end{figure}

\begin{figure}
\centering
\includegraphics[,width = 0.45\textwidth]{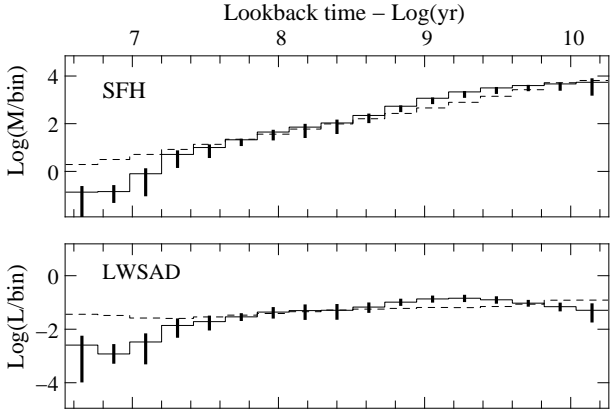}     
\caption{Non parametric photometric inversion ($\alpha$ = 1) of the inner region of NGC 4388. Top and bottom panel: star formation history (SFH) and luminosity-weighted stellar age distribution (LWSAD) vs. lookback time. For a detailed description see Fig.~\ref{innersfr01}.}	
\label{innersfr02}
\end{figure}

\item Photometry ($\alpha = 1$)

The fit reproduces the observations well with a $\chi^2=0.36$ (middle panel of Fig. \ref{innerfit}).
The star formation history (Fig.~\ref{innersfr02}) is quite similar to the star formation history recovered in the case $\alpha = 0$, except in the past 10 Myr. The metallicity is consistent with the spectroscopic results with larger error bars. From the photometric inversion we recovered the average reddening of the stars in the region, $E(B-V) = 0.18$.

\item VLT spectrum + photometry ($\alpha = 0.5$)

The fit of the spectrum is essentially identical to the case $\alpha = 0$, but we do not show it. The photometry is shown in the bottom panel of Fig.~\ref{innerfit}. The total $\chi^2_{\rm{tot}} = 1.4$ is intermediate between the two former cases. The star formation history and the luminosity-weighted stellar age distribution are shown in Fig.~\ref{sfrfit05}.

\end{itemize}

\begin{figure}
\centering
\includegraphics[width = 0.45\textwidth]{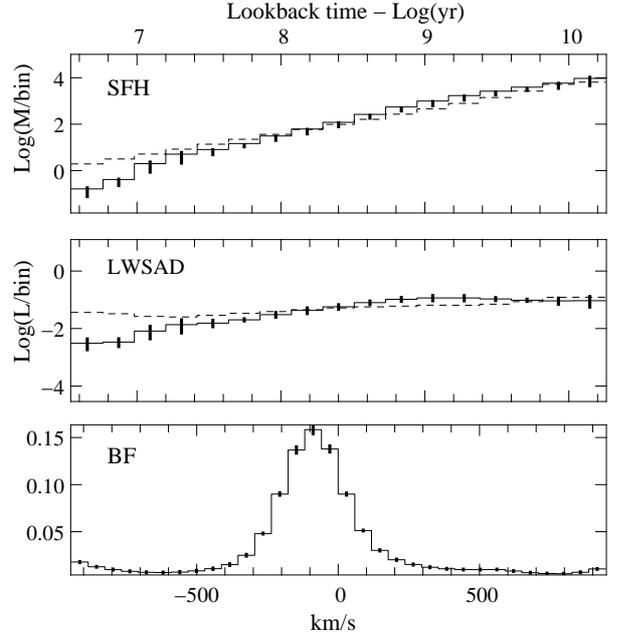}
\caption{Non parametric combined inversion ($\alpha$ = 0.5) of the inner region of NGC 4388. Top and middle panels: star formation history (SFH) and luminosity-weighted stellar age distribution (LWSAD) vs. lookback time. Bottom panel: spectral broadening function (BF). For a detailed description see Fig.~\ref{innersfr01}.}
\label{sfrfit05}
\end{figure}

\subsubsection{Outer region}

\begin{figure*}
   \centering
    \includegraphics[angle = -90,width = 0.9\textwidth]{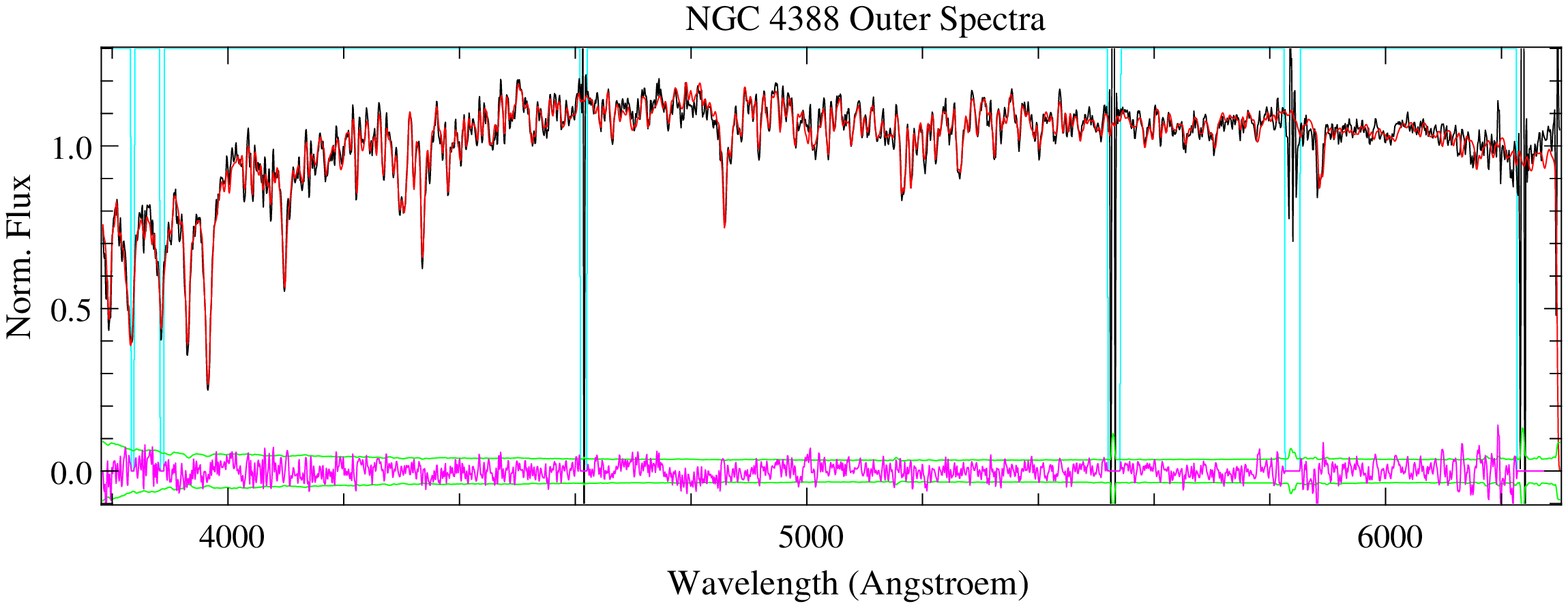}
    \includegraphics[angle = -90,width = 0.9\textwidth]{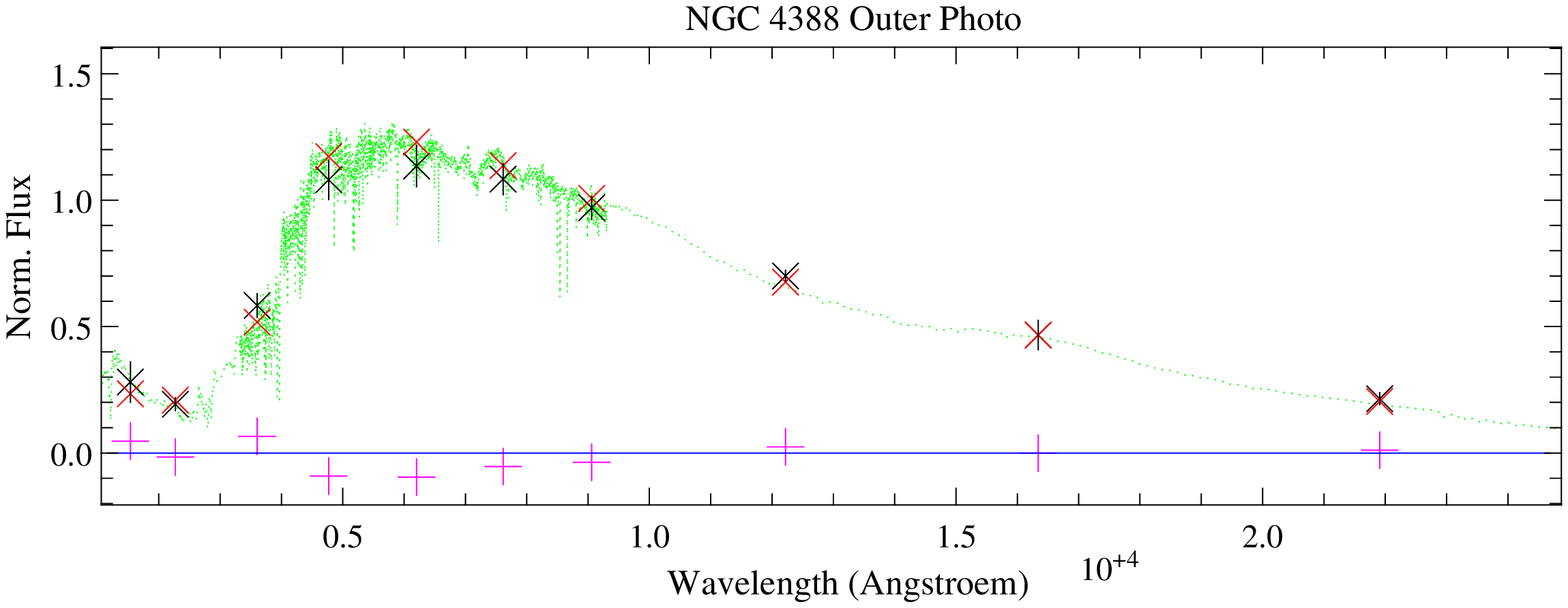}
    \includegraphics[angle = -90,width = 0.9\textwidth]{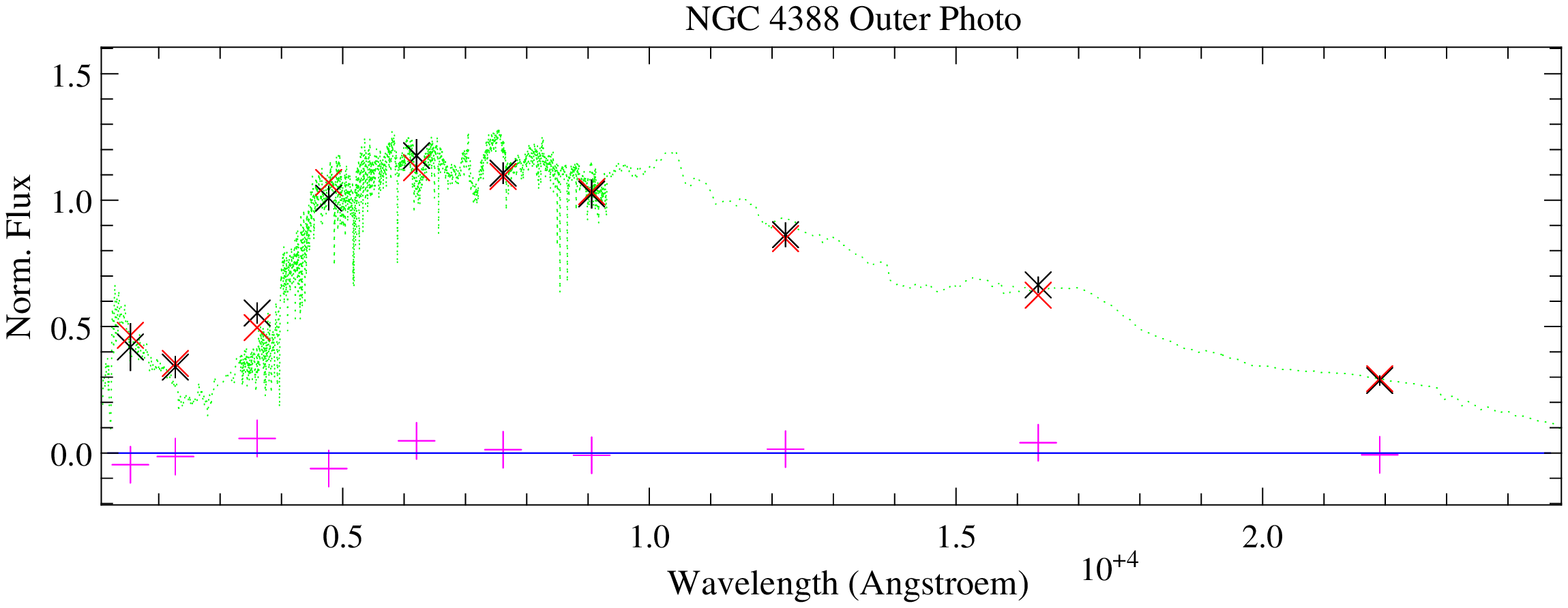}
   \caption{NGC 4388 outer region fits: top panel: spectra (black) and best fit (red) obtained from inversion with $\alpha = 0$. The bottom of the panel shows the residuals (magenta) and the observational errors (green). The cyan vertical lines show the mask used for emission and sky lines. Middle panel: photometry of NGC 4388 (black crosses and error bars) and best fit (red crosses) obtained from inversion with $\alpha  = 1$. The best model spectrum is overplotted. The photometric residuals are also shown (magenta crosses). Bottom panel: same as the middle panel, but using $\alpha  = 0.5$. We do not shown the fit of the spectrum for the case $\alpha =0.5$, because it is indistinguishable by eye from the top panel.}
   \label{outerfit}
   \end{figure*}

\begin{figure}
   \centering
    \includegraphics[,width = 0.45\textwidth]{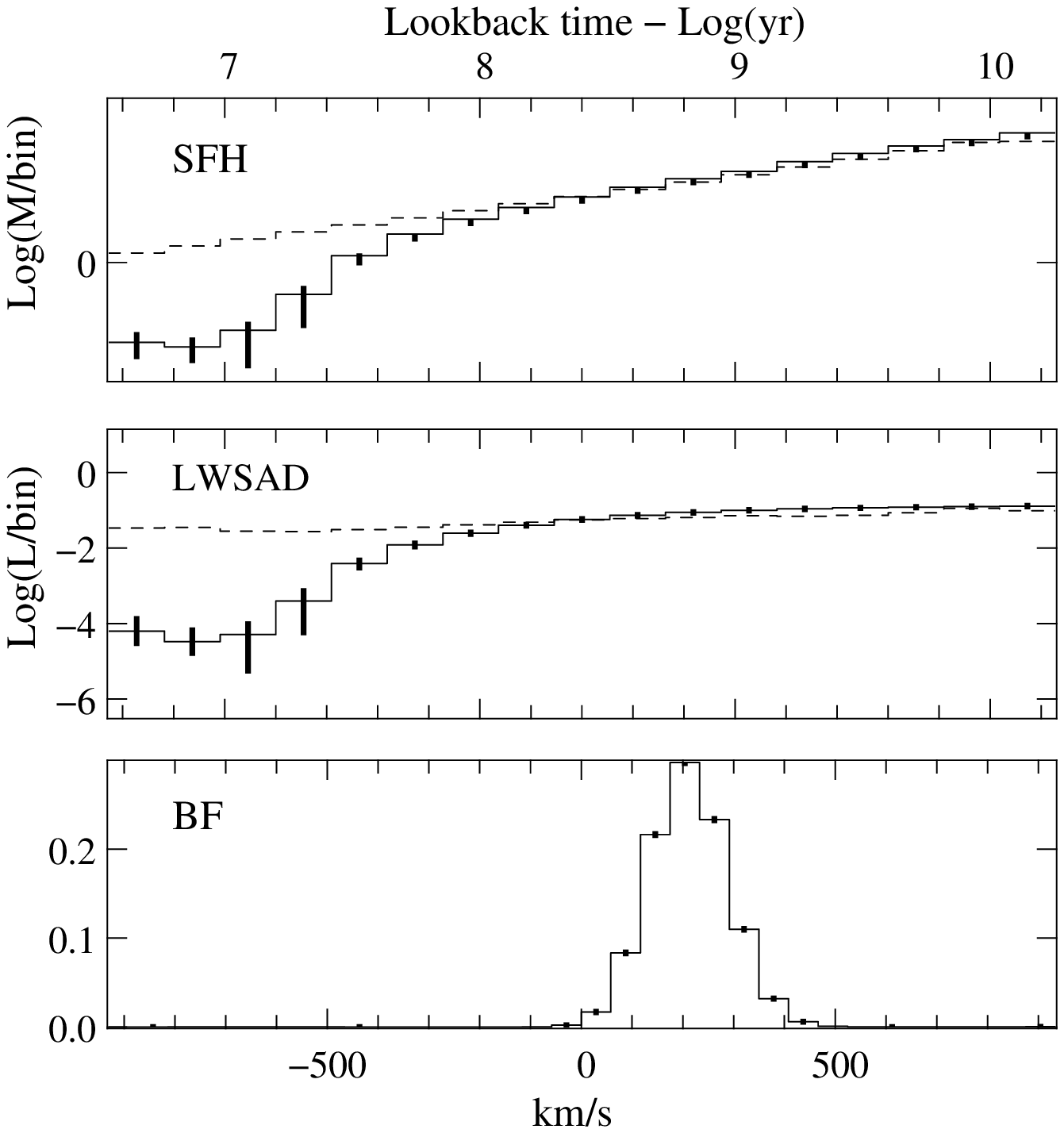}
   \caption{Non parametric spectral inversion ($\alpha$ = 0) of the outer region of NGC 4388. Top and middle panels: star formation history (SFH) and luminosity-weighted stellar age distribution (LWSAD) vs. lookback time. Bottom panel: spectral broadening function (BF). For a detailed description see Fig.~\ref{innersfr01}.}
 \label{outersfr01}
   \end{figure}

\begin{figure}
   \centering
    \includegraphics[,width = 0.45\textwidth]{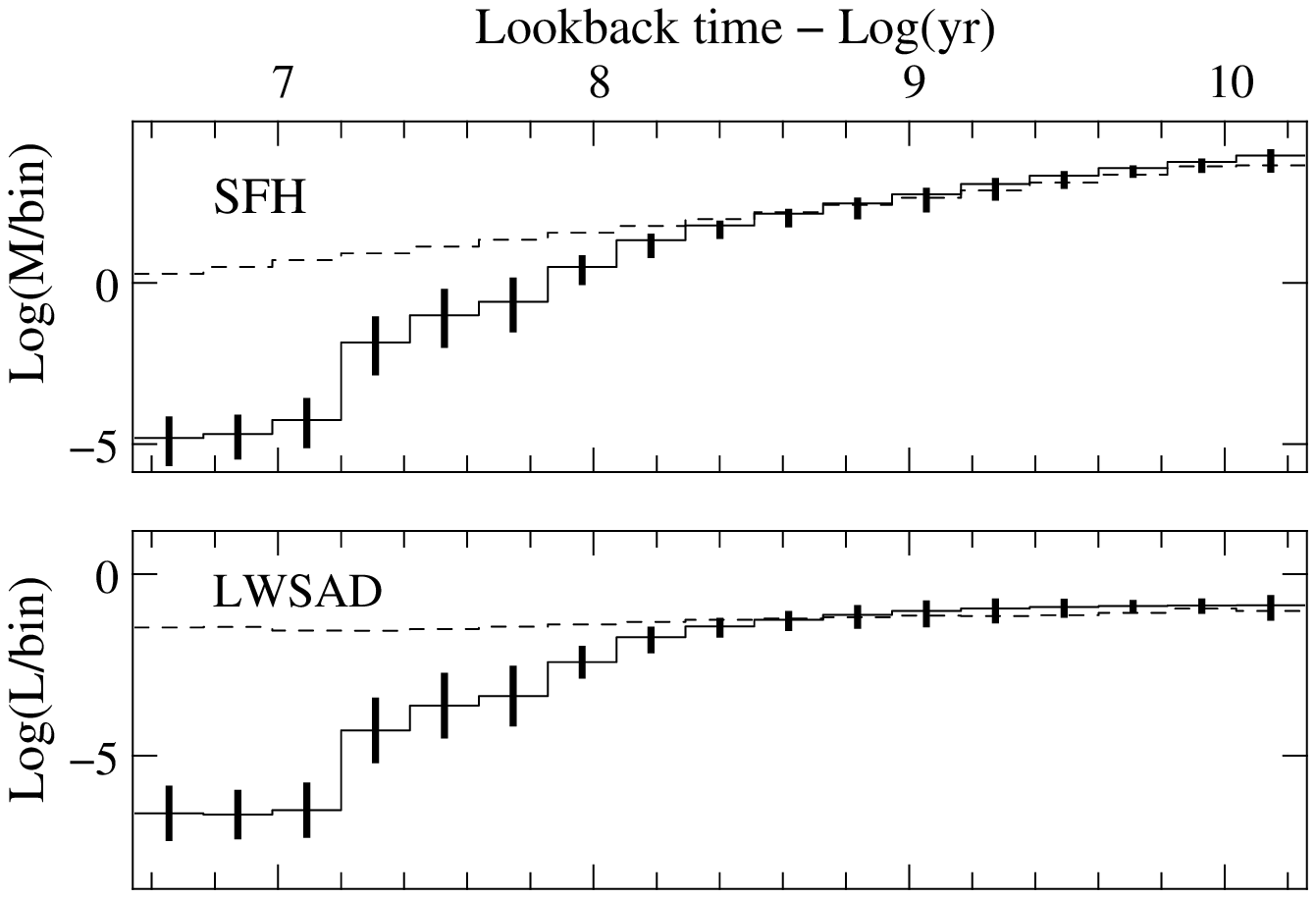}
   \caption{Non parametric photometric inversion ($\alpha$ = 1) of the outer region of NGC 4388. Top and bottom panels: star formation history (SFH) and luminosity-weighted stellar age distribution (LWSAD) vs. lookback time. For a detailed description see Fig.~\ref{innersfr01}.}
 \label{outersfr02}
   \end{figure}

\begin{figure}
   \centering
   \includegraphics[width = 0.45\textwidth]{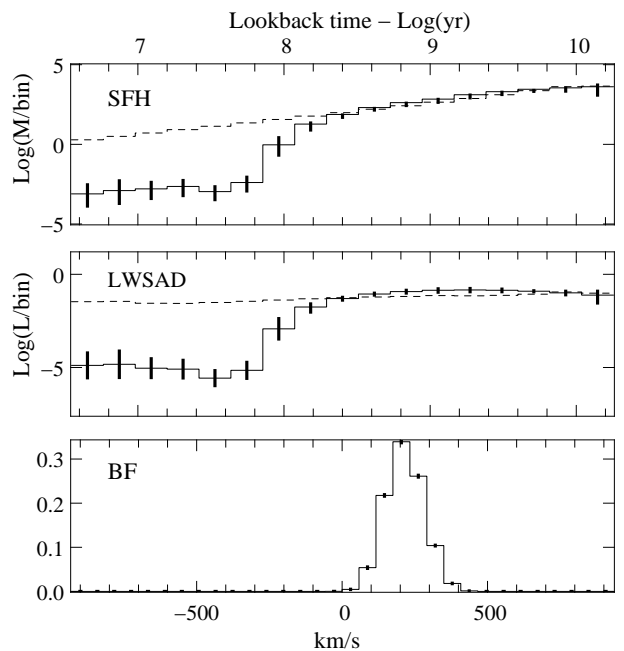}
  \caption{Non parametric combined inversion ($\alpha$ = 0.5) of the outer region of NGC 4388. Top and middle panels: star formation history (SFH) and luminosity-weighted stellar age distribution (LWSAD) vs. lookback time. Bottom panel: spectral broadening function (BF). For a detailed description see Fig.~\ref{innersfr01}.}
  \label{ptmfit05out}
   \end{figure}

\begin{itemize}

\item VLT spectrum ($\alpha = 0)$

We obtain a fit (top panel of Fig.~\ref{outerfit}) with a $\chi^2_{\rm{spec}} = 0.43$. The recovered star formation history (Fig.~\ref{outersfr01}) is flat until a lookback time of $\sim$ 200 Myr at most, then it shows a drop, as expected from the ram pressure stripping scenario.
The time-bin averaged metallicity is $\langle Z \rangle = 0.01 \pm 0.006$, about half of that of the inner region. The broadening function is centered on $\approx$ 200 km/s, consistent with galactic rotation (\citealt{vi2}). Its width is dominated by the spectral broadening of the instrument.

\item{Photometry ($\alpha = 1$)}

The star formation history shows a shape similar to the $\alpha = 0$ case, with a departure from a constant value at a lookback time of $\sim 200$ Myr at most (Fig.~\ref{outersfr02}). The extinction is $E(B-V) = 0.07$ and the $\chi^2_{\rm{phot}} = 1.1$ (middle panel of Fig.~\ref{outerfit}).

\item VLT spectrum + Photometry ($\alpha  = 0.5$)

The total fit (bottom panel of Fig.~\ref{outerfit}) has a $\chi^2_{\rm{tot}} = 0.51$. The spectral fit is very similar to the case $\alpha  = 0$ so is not shown.
The total star formation history was flat until {\bf $\sim 200$} Myr ago, then it decreases steeply (Fig.~\ref{ptmfit05out}). The gas stripping of NGC~4388 truncated the star formation history between 100 and 500 Myr ago.

Fig.~\ref{sfrfit05}-\ref{ptmfit05out} compare the recovered star formation history for the inner and the outer regions to a flat star formation. While the inner region shows good agreement, the outer region, somewhere between 100-200 Myr, significantly deviates from a flat star formation history. Since we regularize the inverse problem by requiring the solution to be smooth, instantaneous truncation is rejected by construction. Therefore we cannot assess whether the progressive nature of the decline of the post-stripping star formation is real or if it is a consequence of the penalization. We get back to this point in Sect.~\ref{62}. 
\end{itemize}

As expected, the non parametric method

\begin{enumerate}
\item provides constraints on the long-term star formation history of the galaxy,
\item confirms a recent radical change in the star formation history
of the outer disk,
\item does not provide a sharply truncated star formation history. 
\end{enumerate}

\subsection{Parametric inversion}

We applied the parametric method to the spectrum of the outer region of NGC 4388 to quantify the time elapsed since star formation dropped by a factor of two from its pre-stripping value, which we call stripping age. For the truncation of star formation, we used star formation rates that decline linearly during 0 to 100~Myr. For the pre-stripping star formation history and metallicity, we used the results of Sect. \ref{nonpmethod}. This is essential because, as explained below, the derived stripping age depends on the luminosity-weighted $Z$ and on the ratio of young-to-old stars at the time of stripping.
The first point to clarify is the importance of the choice of metallicity evolution in the stripping age determination.

To test the influence of different AMRs on determination of the stripping age, we used a grid of constant metallicity models, as well as two AMRs from the galaxy evolution models of \cite{bp}. We find that the stripping age decreases when the average metallicity increases. The stripping age of the parametric method is only sensitive to the metallicity averaged over the past 5~Gyr. Since the time-bin averaged metallicity from the non-parametric method, which is averaged over a Hubble time, is lower by $\Delta Z \sim 0.005$-$0.01$ (cf. Sect.~\ref{nonpmethod}), we took a constant metallicity of $Z = 0.018$ for the parametric method. We studied the $\chi^2$ as a function of the stripping age. As done for the non parametric method, we applied the parametric method to the VLT spectrum, the photometry alone and then spectrum and photometry together. We find that the stripping age is independent of the duration of the star formation truncation. The $\chi^2$ analysis does not allow different durations of the star formation truncation to be distinguished. In the following we only discuss the case of instantaneous truncation.

\begin{itemize}

\item VLT spectrum ($\alpha = 0$)

The top panel of Fig.~\ref{exptau} shows the value of $\chi^2$ as a function of the stripping age, yielding a most likely stripping age of 190 Myr. 

\begin{figure}
\begin{center}
\includegraphics[width = 0.5\textwidth]{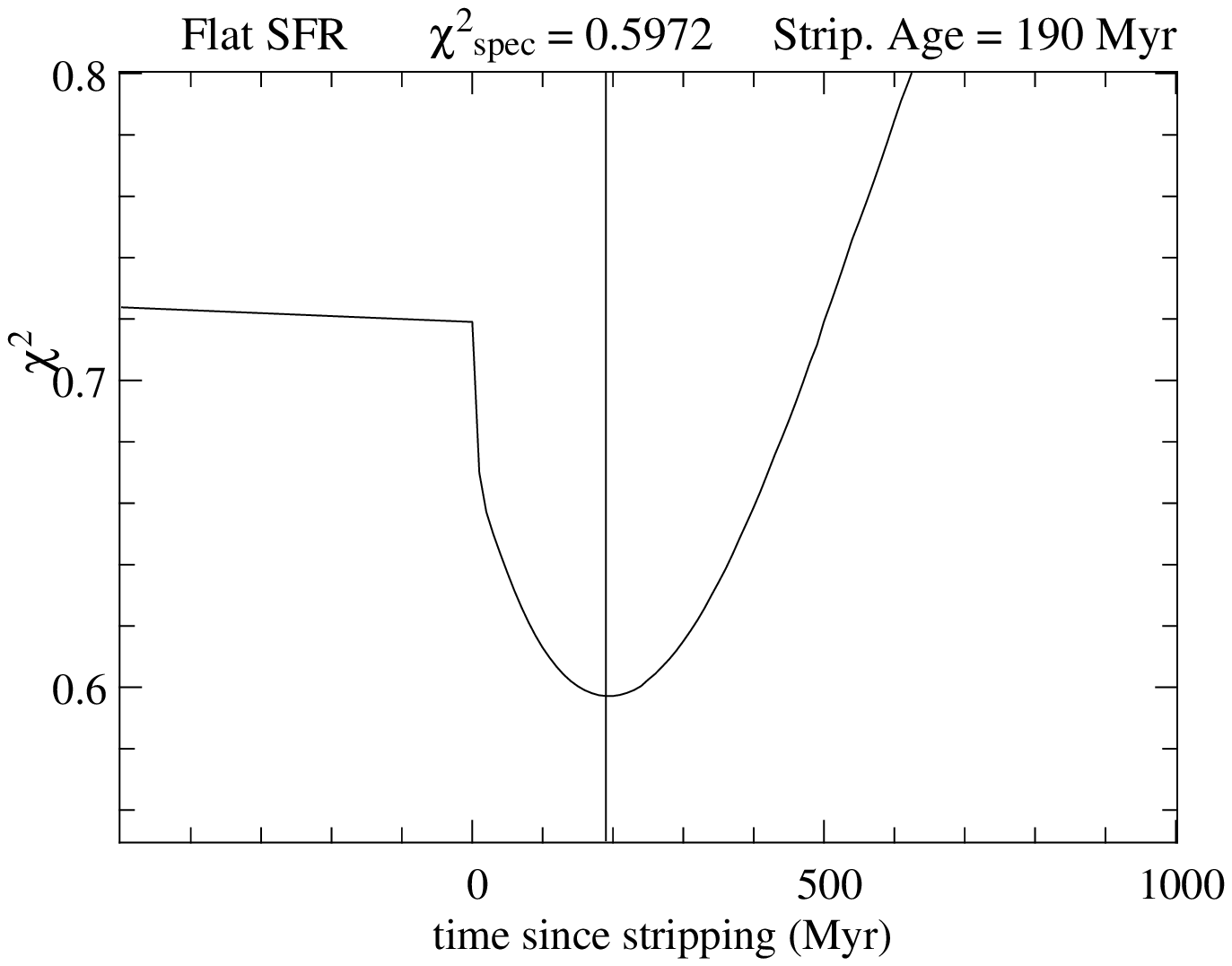}
\includegraphics[width = 0.5\textwidth]{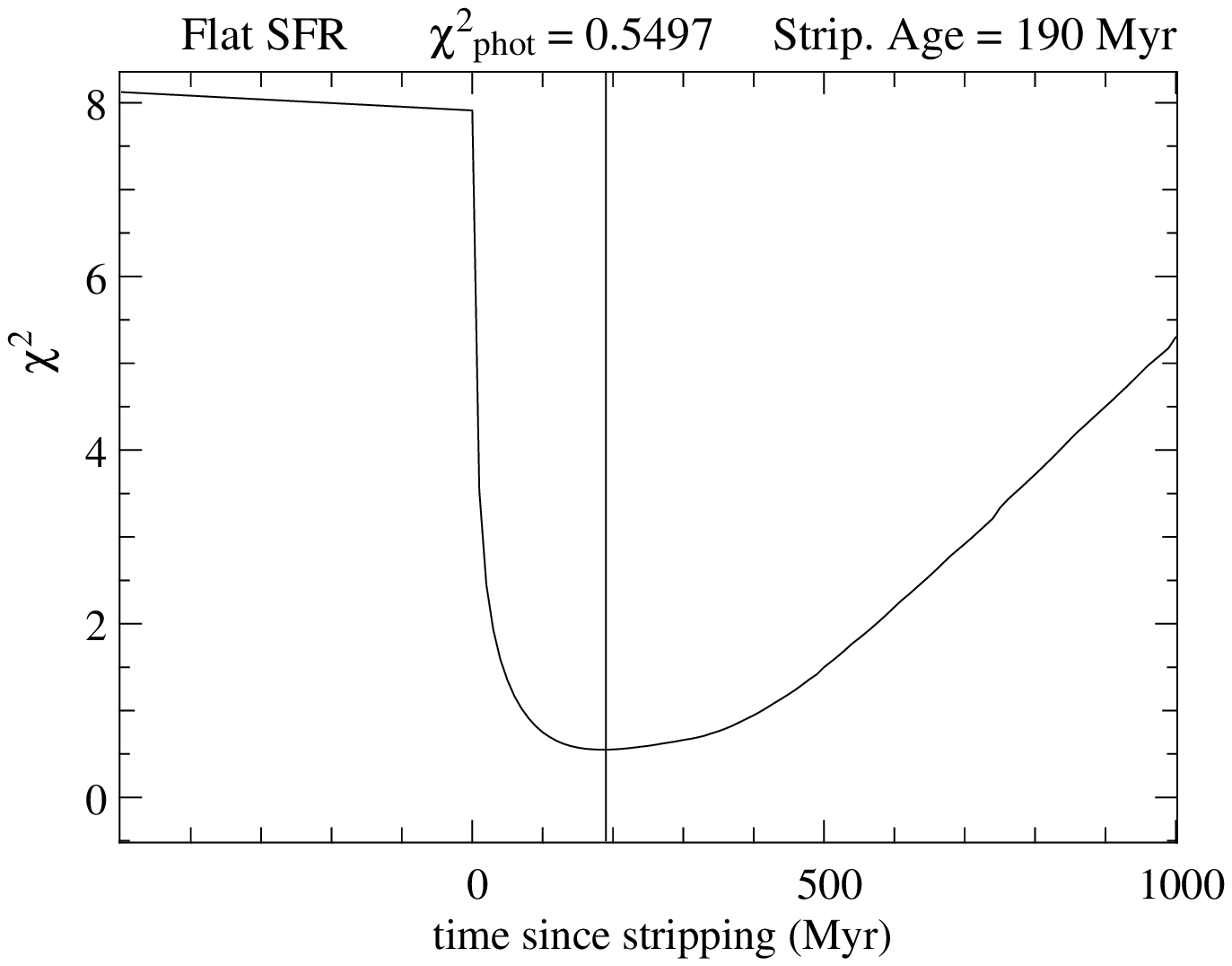}
\includegraphics[width = 0.5\textwidth]{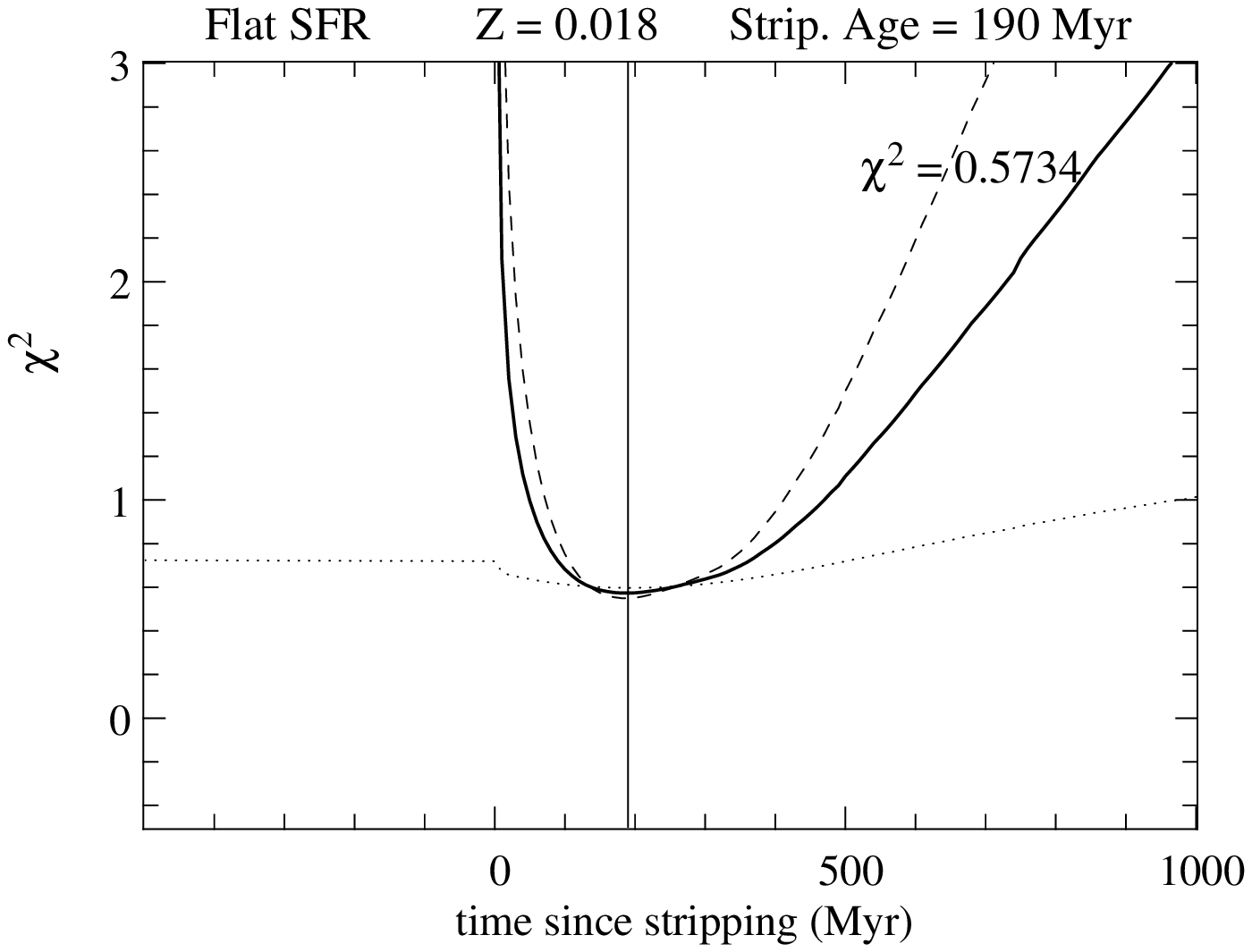}
\caption{Parametric inversion of the outer region for an instantaneous star formation truncation. Each panel shows the $\chi^2$ evolution as a function of the time elapsed since the truncation of star formation. To build the set of spectra, we used a constant star formation rate and metallicity ($Z = 0.018$). Top, middle, and bottom panels show the $\chi^2$ evolution for spectral, photometric, and combined analysis, respectively. The bottom panel overplots the results of the spectral (dotted line) and photometric (dashed line) analysis. Labels indicate the underlying star formation history, the metallicity, the values of the minimal $\chi^2$, and the resulting stripping ages.}
\label{exptau}
\end{center}
\end{figure} 

\item Photometry ($\alpha = 1$)

We obtained a $\chi^2_{\rm{phot}} = 0.54$ (middle panel of Fig.~\ref{exptau}) and a stripping age of 190 Myr consistent with the result obtained for the VLT spectrum analysis. We also recovered an $E(B-V) =  0.1$, in good agreement  with the reddening obtained in the equivalent non parametric problem.

\item VLT spectrum + photometry ($\alpha  = 0.5$)

The minimum $\chi_{\rm{tot}}^2$ is 0.57, and the stripping age is again of 190 Myr (bottom panel of Fig.~\ref{exptau}).

\end{itemize}

With the non parametric method, we fixed an upper limit for the stripping age of about 100-200~Myr. The parametric method improved the precision by giving a stripping age of 190 Myr. It is remarkable that the photometric and spectroscopic data provide good agreement on the stripping age. Uncertainties on this value are discussed in the next section.


\section{Discussion}
\label{discussion}

\subsection{Uncertainties in the parametric method}

In determining of the stripping age, a crucial point is to determine the uncertainty of the result. To clarify this point, we discuss here the influence of the potential sources of errors in the method.

\begin{itemize}

\item[-] {\bf Monte Carlo simulations} 

We add a Gaussian noise to the spectral and photometric data, and perform 500 Monte Carlo simulations that way. As expected
from statistics, the absolute values of the minimum $\chi^2$ vary and the dispersion is $\sim 1/\sqrt{N}$, 
where $N$ is the number of degrees of freedom ($> 1700$ for the spectroscopic analysis, 7 for the photometric analysis). 
However, the location of the minimum in the $\chi^2$ curves varies very little.  
The error on the stripping age associated with the noise is $20$ Myr.

\item[-]{\bf Extinction law}

The extinction law of \cite{cal} was designed to provide a reasonable correction for starburst galaxies. Compared to
other laws in the literature, such as the standard Milky Way extinction laws, it is relatively flat. The choice of the
extinction law affects the UV-optical colors particularly strongly. However, in the external regions of NGC\,4388 from which
gas has mostly been removed by ram pressure stripping, the effect of the choice of the extinction law can only be small.
As a test, we apply to the model spectra two different extinction laws (\citealt{cal}, \citealt{car}), and we compare the recovered stripping ages. The results differ by $10$ Myr.   

\item[-]{\bf Long-term star formation history}

At fixed metallicity, we build families of model spectra with different star formation histories. We consider exponentially 
decreasing star formation rates (SFR  $\propto e^{-\tau/t}$) with timescales $\tau = $6, 10, and 50 Gyr, as well as a 
flat SFR. The stripping age increases from 100\,Myr for $\tau=6$\,Gyr to 190\,Myr for a flat SFR, 
because the ratio of young-to-old stars before the ram pressure stripping event is higher with flatter star formation rates. 
Shorter timescales are excluded, because the corresponding spectra are too heavily dominated by old stars even before
any cut in the SFR is considered. In our study the long-term star formation history is 
constrained strongly by the non parametric inversion of Sect.\,\ref{nonpmethod}. 
For a wide range of penalization weights, constant star formation rates are favored.
The uncertainty on the stripping age obtained when considering only regular star formation histories consistent 
with the non parametric results is reduced to about 10\,Myr.

\item{\bf Metallicity}

At fixed star formation, we build sets of model spectra using different metallicity values. 
The stripping age increases with increasing metallicity. 
With our study the time-bin averaged metallicity over the last 5~Gyr is established through the non-parametric analysis. The error on the metallicity is $20 \%$. The resulting error on the stripping age is $10$\,Myr. 
It is worth noting that the agreement between the stripping ages determined from the photometry alone and 
from the spectroscopy alone is best with the flat star formation history and the quasi-solar metallicity
found by the non-parametric analysis. With other assumptions, this agreement is lost. 
\end{itemize}

\subsection{Gas dynamics and star formation}
\label{62}

The non parametric method fixed an upper limit for the stripping age of about 100-200 Myr, with a rather unconstrained duration for the truncation. 
Numerical simulations (Abadi et al. 1999, Vollmer et al. 2001, R\"{o}diger \& Br\"{u}ggen 2006) show that the ISM located at a given distance from the galaxy center is stripped rapidly within a few $10$~Myr. Therefore in the family of possible interpretations of the outer disk data, we favor a posteriori the scenarios with sharp truncations, although the timescale of the decline cannot be constrained with the data presented here. The instantaneous truncation scenario can only be studied with a parametric method, since the penalization used in the non-parametric method rejects luminosity weighted stellar age distributions with sharp variations. A rapid halt in star formation suggested by 
the dynamical models can only be studied with a parametric method, where we make the simplifying assumption that the gas is stripped
instantaneously. The derived stripping age of $190$~Myr is consistent with the extent of the observed H{\sc i} tail (Oosterloo \& van Gorkom 2005):
a tail extent of $80$~kpc in the plane of the sky and along the line of sight with the given stripping age leads to a total velocity of $\sim 570$~km\,s$^{-1}$ and a radial velocity of $\sim 400$~km\,s$^{-1}$. The latter corresponds to the
difference between the galaxy's systemic velocity and the velocity of the H{\sc i} tail (Oosterloo \& van Gorkom 2005). The full extent of the H{\sc i} tail was not known at the time when Vollmer \& Huchtmeier (2003) presented their dynamical model. Their tail has an extent of $40$~kpc
with an associated stripping age of $120$~Myr. A new, revised dynamical model (Vollmer, in prep.) with the observed extent yields a ram pressure peak $\sim 170$~Myr ago, consistent with our findings.

\section{Conclusions}
\label{conclusion}

VLT FORS2 spectroscopic observations of the inner star-forming and outer gas-free disk of the Virgo spiral galaxy NGC 4388 have been presented. 
Previous observations indicate that this galaxy has undergone a recent ram pressure stripping event. Once the galaxy's ISM is stripped by ram
pressure star formation stops. We detect this star formation truncation in the spectrum and the multiwavelength photometry of the outer disk region of NGC 4388.

To derive star formation histories we extended the non parametric inversion method of \cite{oc} making a joint analysis of spectroscopic and photometric data possible. The new code was tested on a series of mock data using Monte Carlo simulations. We find that
the results are stable, once that minimization has converged. The uncertainties for the young stellar ages ($\le 10$ Myr) can be large because of the uncertainties 
of stellar models. Whitin reasonable limits, e.g. acceptable $\chi^2$, the shape and the normalization of the initial guess does not significantly affect the recovered solution (at a fixed signal-to-noise ratio).

The new inversion tool was applied to our spectroscopic and photometric data for NGC 4388. We explored the effect of different penalizations in case of spectral analysis, photometric analysis and combined analysis. The main results are (i) the recovered star formation history is flat for the inner disk spectrum and (ii) for the outer disk region the inversion yields a star formation drop at a lookback time of $\sim$ 200 Myr at most.

We have introduced a parametric method that refines the determination of the stripping age, the time elapsed since the star formation has dropped by a factor of two from its pre-stripping value. Based on the non parametric results, we assumed a flat star formation before the stripping event and an almost solar metallicity. We approximated the effect of gas stripping by cutting or decreasing linearly the star formation at a different lookback time $0 \le t \le 1$ Gyr. The obtained set of spectra was compared with the observed spectrum of the outer disk of NGC 4388. 
The effect of the potential sources of error in the stripping-age determination were evaluated.

The parametric method leads to a stripping age for NGC 4388 of $\sim 190 \pm 30$ Myr. It cannot distinguish between an instantaneous and a longer lasting star formation truncation. The derived stripping age agrees with the results of a previous work of \cite{cr} and with revised dynamical models.

\begin{acknowledgements}
This publication makes use of data products from the Two Micron All Sky Survey, which is a joint project of the University of Massachusetts and the Infrared Processing and Analysis Center/California Institute of Technology, funded by the National Aeronautics and Space Administration and the National Science Foundation.

Funding for the SDSS and SDSS-II has been provided by the Alfred P. Sloan Foundation, the Participating Institutions, the National Science Foundation, the U.S. Department of Energy, the National Aeronautics and Space Administration, the Japanese Monbukagakusho, the Max Planck Society, and the Higher Education Funding Council for England. The SDSS Web Site is http://www.sdss.org/.

We would like to thank D.~Munro for
freely distributing his Yorick programming language (available at
\texttt{http://www.maumae.net/yorick/doc/index.html}). We thank the referee for the useful comments.
   \end{acknowledgements}


\end{document}